\documentclass[aps,prl,letterpaper,11pt,twoside,tightenlines,nofootinbib,showpacs,preprint,twocolumn]{revtex4}
\usepackage{graphicx}
\usepackage[sort&compress]{natbib}
\usepackage{latexsym}
\usepackage{epsfig}
\begin{document}
\newcommand{\eg}{{\it e.g.}}
\newcommand{\etal}{{\it et. al.}}
\newcommand{\ie}{{\it i.e.}}
\newcommand{\be}{\begin{equation}}
\newcommand{\dd}{\displaystyle}
\newcommand{\ee}{\end{equation}}
\newcommand{\bea}{\begin{eqnarray}}
\newcommand{\eea}{\end{eqnarray}}
\newcommand{\bef}{\begin{figure}}
\newcommand{\eef}{\end{figure}}
\newcommand{\bce}{\begin{center}}
\newcommand{\ece}{\end{center}}
\def\lsim{\mathrel{\rlap{\lower4pt\hbox{\hskip1pt$\sim$}}
    \raise1pt\hbox{$<$}}}         
\def\gsim{\mathrel{\rlap{\lower4pt\hbox{\hskip1pt$\sim$}}
    \raise1pt\hbox{$>$}}}         

\title{A  reanalysis of Finite Temperature SU(N) Gauge Theory}
\author{P.~Castorina${}^{1,2}$, V.~Greco${}^{1,3}$, D.~Jaccarino${}^{1,3,4}$, D.~Zappal\`a${}^2$}

\affiliation{
\mbox{${}^1$ Dipartimento di Fisica, Universit\`a di Catania, Via Santa Sofia 64, I-95123 Catania, Italy.}\\
\mbox{${}^2$ INFN, Sezione di Catania, I-95123 Catania, Via Santa Sofia 64, Italy.}\\
\mbox{${}^3$  INFN, Laboratori Nazionali del Sud, I-95123 Catania, Italy.}\\
\mbox{${}^4$ Scuola Superiore di Catania - Via Valdisavoia 9 I-95123 Catania, Italia }
}

\date{\today}
\begin{abstract}
\vskip10pt

We revise the $SU(N_c)$, $N_c=3,4,6$, lattice data on  pure gauge theories at
finite temperature by means of a quasi-particle approach.
In particular we focus on the relation between the quasi-particle effective mass and the order of the deconfinement
transition, the scaling of the interaction measure with $N^2_c -1$,  the role of gluon condensate, the screening mass.
\end{abstract}
\maketitle

\section{I. Introduction}

A careful investigation of the quark-gluon plasma phase needs an understanding of the details of the deconfinement
transition which occurs above a critical temperature $T_c$ and in this respect the data provided by lattice simulations
represent the best tool for testing various dynamical models  close to $T_c$. Besides the full QCD simulations,
interesting indications on the gluonic  sector can be drawn from lattice studies on the high temperature transition
of pure non-abelian gauge theories $SU(N_c)$, where all difficulties related to the presence of fermions and
to the details of the chiral simmetry breaking are absent.  Recent lattice simulations on $SU(N_c)$  gauge  theories
at finite temperature $T$ and for large number of colors $N_c$ \cite{panero,gupta1} are now available
and they show a peculiar scaling of the interaction measure, $\Delta=(\epsilon-3p)/T^4$
( $\epsilon =$energy density and $p=$pressure), with the number of gluons $N^2_c-1$.
Moreover, in the range $1.1 \phantom{.}T_c < T < 4 \phantom{.} T_c$, the interaction measure has a $O(1/T^2)
$ behavior which implies a $O(T^2)$ contribution in $\epsilon-3p$  \cite{pisarski}.

The previous features have interesting consequences on the number of effective degrees of freedom and on the role of
the gluon condensate above the critical temperature. The observed approximate scaling of $\Delta$ with  $N^2_c-1$ for
$N_c \ge 3$,  suggests: 1) a quasi-particle behavior of the effective degrees of freedom, with the typical
 degeneracy, $N^2_c-1$, of the gluons  and  with an effective, temperature dependent, mass  that turns out to be
divergent or, at least, very large at $T_c$ \cite{peshier,mannarelli,redlich,Plumari:2011mk};
2) the presence of the same degeneracy factor in the gluon condensate contribution, if any,  to $\Delta$;
3) the same proportionality to the number of gluons $N^2_c-1$ of the dynamical mechanism which produces the $O(T^2)$
contribution.

However, a more accurate analysis shows that  the scaling of interaction measure and of the
other thermodynamic observables with $N^2_c-1$ is not exact \cite{gupta1}.
More precisely one can consider two different ranges of temperature. Near the transition temperature , i.e.
for $T_c<T<(1.05-1.1) T_c$, the scaling with $N_c^2-1$ is clearly violated as shown in Fig.1, whereas above
the $\Delta$ peak temperature  the scaling is  almost exact , see Fig.2.

In this brief report we reconsider  the $SU(3,4,6)$ lattice data 
in \cite{boyd,gupta1}, by
means of  a quasi-particle approach in order to discuss  the origin
of the deviations from the exact scaling in relation to the critical
behavior. The general formalism is discussed Sec. II; the comparison
with lattice data is presented in Sec. III;
the role of the gluon condensate is
addressed in Sec. IV and in Sec. V we draw our conclusions.

\begin{figure}
\epsfig{file=deltanearTcdata.eps,width=8.0cm}
\caption{$\Delta/(N^2-1)$ close to  $T_c$.}
\end{figure}

\begin{figure}
\epsfig{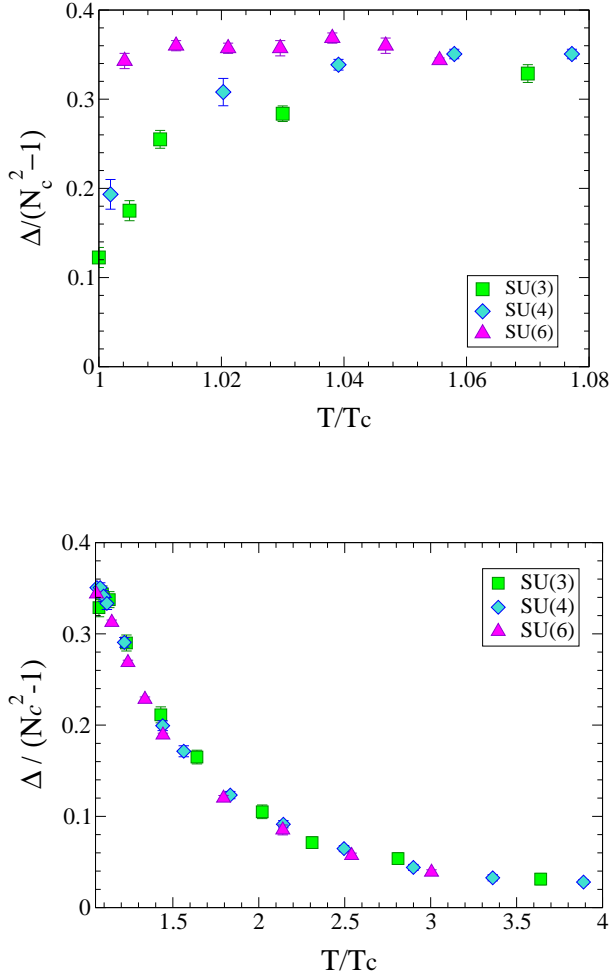}
\caption{$\Delta/(N^2-1)$ versus $T/T_c$ above the $\Delta$ peak.}
\end{figure}

\vskip 20pt

\section{II. General formalism}

The partition function for the very simple case of free quasi-particles
in a volume $V$, at temperature $T$ and with temperature dependent mass $m(T)$ is
\bea
\label{partitionfunct}
& {\rm ln} {\cal Z }(T,V) =  2V (N_c^2-1) \times \nonumber\\
& {\displaystyle \int \frac{d^3k}{(2\pi)^3} } ~ {\rm ln} \left[ f_T(k) ~ {\rm exp} \left
( \sqrt{\vec k^2 + m^2(T)} / T \right)\right] \eea 
where $f_T(k)$ is the distribution
\be
\label{bedistribution}
 f_T(k)=\left [ {\rm exp} \left (\sqrt{\vec k^2 + m^2(T)} / T  \right )  -1 \right ]^{-1} 
\ee
and all
thermodynamical quantities are obtained by deriving Eq.
(\ref{partitionfunct}). For our purpose, the energy density
$\epsilon$, the pressure $p$, the entropy density $s$ are
respectively, $\epsilon= (T^2/V) ~\partial \ln {\cal Z} / \partial
T$ ,  $p= T ~ \partial \ln {\cal Z} / \partial V$ and $s=(\epsilon +
p)/T$. The interaction measure $\Delta$ is directly obtained from
$\epsilon$ and $p$ as $\Delta=(\epsilon-3p)/T^4$ .

Obviously the temperature dependence of the mass must be taken into account
when differentiating with respect to $T$. Note also that the additional effect
of a temperature independent bag pressure
(gluon condensate) $B$ corresponds to the  changes
$p \rightarrow p - B$, and $\epsilon \rightarrow \epsilon + B$,
with no change in the entropy density $s$.

The factor 2  in front of the color multiplicity factor $N_c^2-1$ in
Eq. (\ref{partitionfunct}), corresponds to the number of
polarization degrees of freedom. In general, the representations of
the Poincar\'e group of massive and massless particles in this case
would suggest that our massive physical constituents  carry three,
rather than two, spin degrees of freedom.  However, this is valid
for free particles and, in fact,  the comparison of all predicted
thermodynamic quantities with the observed high temperature lattice
results clearly shows much better agreement when only two
polarization states are considered. 

In particular all lattice QCD results, as shown also in Fig.5,
hint at an asymptotic limit of the $\epsilon/T^4$ consistent with
$2\,(N_c^2-1)$ degrees of freedom.
In other words, a simple shift
to massive gluons with three spin degrees of freedom cannot
satisfactorily explain the effects of the interaction apparently
still present in the gluon gas above the critical temperature.

It is well known that gauge symmetry forbids a mass term in the
lagrangian for the elementary gluons and, in order to preserve the
symmetry, one can expect to observe the generation of mass  through
a dynamical mechanism, such as the Schwinger mechanism \cite{schwin}
in which the mass comes from the appearance of a pole in the
self-energy. In fact this effect has been explicitly pointed out  and
it has been argued that the longitudinal polarization component could 
be canceled by the scalar massless pole \cite{jackiwjohnson,cornwall,papav}.

On the other hand, in the modified Hard Thermal Loop perturbation
theory approach, where each order already includes some
aspects of gluon dressing and which leads to a rather rapid
convergence of the expansion, the contribution of longitudinal
gluons vanishes in the limit $g \to 0$, and, in particular, one also
obtains the right number of degrees of freedom for the
Stefan-Boltzmann form \cite{resum}.

Moreover, from a comparison of the lattice glueball spectrum with
the predictions of constituent models it has recently been argued
\cite{Mathieu} that massive gluons should in fact be transversely
polarized, since two massless gluons cannot combine to form a
longitudinally polarized massive gluon \cite{yang}. According to
these indications we limit ourselves to consider just to
polarization degrees of freedom for the effective quasi-particle in
Eq. (\ref{partitionfunct}).

Let us now turn to the most important ingredient in our approach,
that is the effective temperature dependent mass $m(T)$ which
contains the non-perturbative dynamics.  Previous analyses
\cite{peshier,mannarelli,redlich} show that $m(T)$ strongly increases near
$T_c$  and a  qualitative explanation of it has been outlined in
\cite{cms}. To illustrate this aspect one describes the mass of the quasi-gluon in
the strongly coupled region as the energy
contained in a region of volume $V_{cor}$ whose characteristic size
is given  by the correlation range $\xi$, so that in three spatial
dimensions one gets ($\eta$ is the anomalous dimension and $t$ is
the reduced temperature $t=T/T_c$):

\be m(t) \simeq \epsilon(t) V_{cor} = \epsilon(t) \int dr~
r^2\frac{\exp[-r/\xi(t)]}{r^{1-\eta}}  \label{mass}\ee

In the case of a second order phase transition, the correlation
length shows the power law divergence $ \xi(t)=(t-1)^{-\nu}$ at
$t=1$, which indicates that the associated fluctuations have an
infinite range at criticality, and the corresponding component of
the energy density vanishes as $\epsilon(t) \simeq
(t-1)^{1-\alpha}$ where $\alpha$ is the specific heat critical
exponent. In this case Eq. (\ref{mass}) predicts a power law
divergence of the mass $m(1)$. Then, by focussing on  the 3D Ising
model universality class, which includes the critical
behavior of the $SU(2)$ gauge theory deconfinement
transition,  one finds the value of the
critical exponent for our mass: $m(t)\simeq (t-1)^{-0.41} $
 In addition the $SU(2)$ critical behavior suggests that very close to
$T_c$, $\Delta$ should have the form \cite{engels1}
 \be \Delta= A\tau (1-\tau)^{0.89} + B \tau \ee
with $\tau=t^{-1}$ and $A,~B$ constants and the exponent is given by $0.89=1-\alpha$.

In such transitions  $m(t)$ close to $t=1$ has a power
law divergent behavior which has to be considered as an approximation
for $T$ near $T_c$. By recalling that for $ t >> 1$, the mass is
expected to grow linearly with the temperature, which is the only
dimensionful scale available, a suitable ansatz for $m(t)$ is 
\be m(t)= \frac{a}{(t-1)^c} + b t , \label{ansatz1}\ee
where $a,b,c$ are constant parameters.

For the gauge groups $SU(N_c)$, with $N_c=3,4,6$ here considered, a
first order phase transition  and consequently a finite correlation
length, is expected at $T_c$ and power law behavior at criticality
is modified by the finite scale $\xi$. However, in the case of  weak
first order transitions  one should  expect a behavior of the
thermodynamical quantities at $T_c$ not totally different from that
observed in second order transitions, and therefore a finite but
large correlation length and a corresponding large $m(1)$.

In particular, as discussed in \cite{repprogphys}, the
thermodynamical quantities approach $T_c$ (from larger values of the
temperature $t>1$) as in a second order phase transition with the
critical temperature shifted to a lower value: $1 \to \delta$ with
$0\sim (1-\delta)<<1$. According to this suggestion the ansatz
(\ref{ansatz1}) for weakly first order phase transition must be
changed into
\be m(t)= \frac{a}{(t-\delta)^c} + b t \label{ansatz2}\ee
which yields a large but finite value of the effective mass and
non-vanishing interaction measure  at the transition point $t=1$.

The quasi-particle mass $m(t)$ should not to be confused with the screening
mass $m_D(T)$. The relation between $m(t)$ and $m_D(T)$, has been clarified in \cite{peshier} 
where it is shown that
\be\label{debyemass}
m_D= \frac{g^2 N_c }{\pi^2 T} 
\int_0^\infty dk ~k^2 f^2_T (k) 
\exp{ \frac{\sqrt{\vec k^2+m(T)^2} }{T} }
\ee
and the leading order QCD coupling $g^2$ is evaluated at the average, $M^2$, 
over the squared quasi-particle momenta, i.e.
\be\label{debye2}
M^2(T) = \frac{4}{3} \frac{\int dk~ k^4 ~f_T(k) }
{\int dk~ k^2 ~f_T(k) }
\ee 
To illustrate this point  in the next Section we display $m(t)$ and $m_D(T)$ 
which show totally different behaviors when approaching $T_c$.

\begin{table}
\begin{center}
\begin{tabular}{ccccc}
 \hline
  \hline
$N_c$ &$c$ & $m(T_c)/T_c$ & $\delta$ & $\chi^2_{dof}$ \\ \hline
$3~~~$& $ 0.5 ~(0.46) ~~$ &$ 6.33 ~(6.64) ~~$ & $0.940 ~(0.952) ~~$ & $2.6 ~(2.0)$ 
\\
$4~~~$& $ 0.5 ~(0.35) ~~$ &$ 6.02 ~(7.68) ~~$ & $0.944 ~(0.982) ~~$ & $7.7 ~(0.8)$ 
\\
$6~~~$& $ 0.5 ~(0.33) ~~$ &$ 4.33 ~(5.91) ~~$ & $0.888 ~(0.969) ~~$ & $6.5 ~(5.8)$ 
\\
 \hline \hline
 \end{tabular}
\end{center}
${}$\vskip-.5cm \caption{\label{table1} Mass, shift $\delta$, and  $\chi^2/dof$
from the fit to SU(3,4,6) lattice data with $a,b,\delta$ as free parameters and $c=0.5$ fixed. 
In  brackets, the same  quantities from the fit with $a,b,c,\delta$  as free parameters.}
\end{table}

\vskip 20pt

\section{III. Analysis of the lattice data}

Now we check the simple model outlined in Sect. II against the
lattice data  for  $SU(3,4,6)$ \cite{boyd,gupta1} . These theories
undergo a  weakly first order  transition and we shall resort to the
large but finite mass in Eq. (\ref{ansatz2}) for computing the
various thermodynamical quantities.

In addition, as discussed above, the singular behavior of the 3D
Ising model corresponds to the value or the critical exponent in Eq.
(\ref{ansatz1}) $c=0.41$, which is close to the mean field exponent
$c=0.5$. Therefore, it seems reasonable to verify whether the mean
field behavior produces a good fit to the data. 

In Table 1 we collect the 
values of the mass $m(1)$, of the shift $\delta$ and of the $\chi^2$
{\it per} degree of freedom (dof), obtained by fitting
the data  with $a,b,\delta$ free parameters and $c=0.5$ fixed.
As an example the $SU(3)$ values of the other parameters $a$ and $b$,
turn out to be $a/T_c= 1.42$ and $b/T_c= 0.53$.
The critical mass decreases when $N_c$ is increased and  
the values of the $\chi^2$ indicate a reasonably good agreement 
with the data.

\begin{figure}
\epsfig{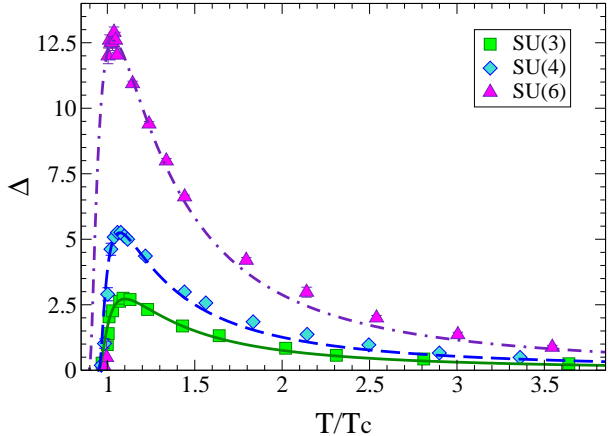}
\caption{The interaction measure as obtained from a fit to 
the $SU(3,4,6)$ lattice data with $c=0.5$ fixed. }
\end{figure}

\begin{figure}
\epsfig{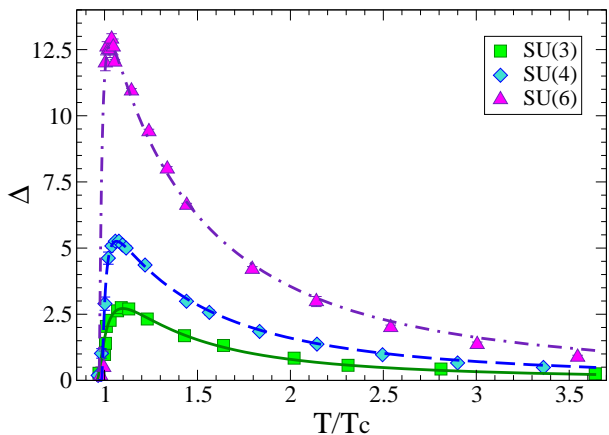}
\caption{The same fit as in Fig.3 but optimizing also on the parameter $c$.}
\end{figure}

\begin{figure}
\epsfig{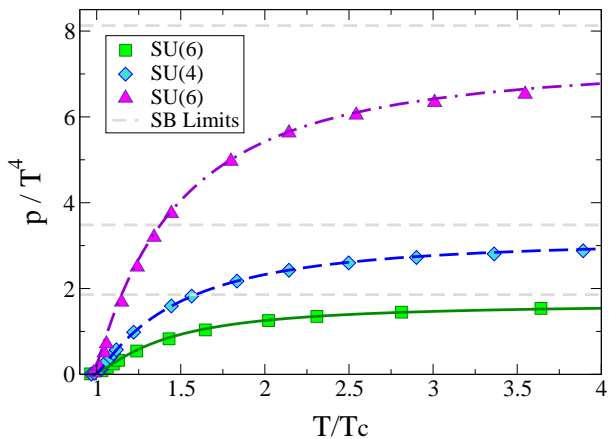}
\caption{Fit to the  pressure of $SU(3,4,6)$ optimizing also on the parameter $c$.}
\end{figure}

However no clear dependence on  $N_c$
can be traced in the values of $m(t)$ and this is expected
because the dependence of the interaction measure $\Delta$,
shown in Figs. 1, 2, on $m(t)$ is highly non-linear.
On the other hand the different behavior of $\Delta$ in Fig. 1, i.e. near the critical temperature,
is essentially due to the dependence of $m(T_c)$ on $N_c$ given in Table I.

It is then remarkable that the simple parametrization
of $m(t)$ in Eq.(\ref{ansatz1},\ref{ansatz2})
produces the correct behavior of the interaction measure $\Delta$
which is plotted in Fig. 3.

An improvement on these results is obviously  obtained by  
releasing the constraint on $c$ and leaving it as another free
parameter of the fit. Results are again reported in Table 1 (in brackets)
and the interaction measure $\Delta$ and the pressure $p$ are plotted
in Figs. 4, 5.  In this case the $\chi^2/dof$ shows an excellent agreement
with the data.

We note that even if we have not forced any specific dependence of the mass 
on the coupling $g$ and on $N_c$ the results of the fit shown in Fig.6 manifests an independence
of $m(T)$ on the the $SU(N_c)$ gauge group for temperature above the peak in the interaction
measure. This could be expected if one
considers the parametric dependence  of the mass in a perturbative approach,
$m_g^2 \simeq g^2 N_c T^2$, along with the t'Hooft scaling of the coupling, 
$g^2 \simeq 1/N_c$.

Moreover, the comparison between the gluon effective mass $m(t)$ and the Debye screeening 
mass $m_D(T)$ according to Eqs. (\ref{debyemass},\ref{debye2}), is displayed in Fig. 6. 
Finally in Fig. 7  $c^2_s=\partial p /\partial \epsilon$ is reported.

As mentioned in the Introduction, lattice results very close to $T_c$ seem to 
scale approximately with $ \simeq N_c(N_c^2-1)$  rather than with $(N_c^2-1)$.  
Accordingly also the behavior of $m(t)$ very close to $T_c$ breaks its independence
on $N_c$, even if not clearly visible from Fig. 6.

This signals that the  t'Hooft condition of QCD at large 
$N_c$, i.e. $g^2 \simeq 1/N_c$,  is violated at the transition region.  
The effective mass at $T_c$  depends on $N_c$ in a way which is not consistent 
with the standard $O(1/N_c^2)$ corrections in QCD at large $N_C$.
This interesting aspect will be reconsidered in the next Section.

\vskip 20pt

\begin{figure}
\epsfig{file=debyemass.eps,width=8.0cm}
\caption{The screening mass $M_D$ compared with $m(t)$ 
as obtained from the fit to the lattice data. }
\end{figure}

\begin{figure}
\epsfig{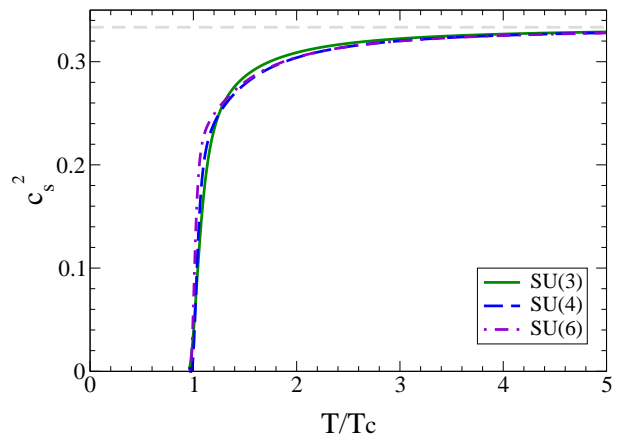}
\caption{The speed of sound as obtained from the fit to the lattice data.}
\end{figure}

\section{IV. The role of the gluon condensate}

A fundamental ingredient of the non perturbative QCD regime is the gluon condensate which has been evaluated by
lattice simulations at zero and finite temperature, in  quenched and unquenched QCD \cite{digiacomo} . It turns out
that:

1) for $T < T_c$ the gluon condensate is almost $T$ independent;

2) for $T> T_c$ the chromo-electric part of the gluon condensate quickly decreases to zero whereas the magnetic one is
constant. This corresponds to the deconfinement transition.

In QCD the gluon condensate is related to the trace anomaly of the energy-momentum tensor by the general expression:
\be
\theta_{\mu}^\mu = 4 B = \frac{\beta(g)}{g} <G_{\mu \nu}^a G_{\mu \nu a} >
\ee
where $B$ is the bag pressure, $\beta(g)$ is the QCD $\beta$-function, $a=1,2,.., (N_c^2-1)$.

Above $T_c$ its contribution to interaction measure is about a half of the zero temperature value ( because the
electric-part melts) and,
as discussed in ref. \cite{cms}, a temperature independent gluon condensate/bag pressure is not able to fit lattice
data and in particular the behavior $\Delta * T^2 \simeq const.$ observed in the range $1.1 T_c < T < 4 T_c$.

To explain this  behavior one has to include in $B$ a term proportional to $T^2$, as already 
suggested in  \cite{pisarski} 
\be\label{tquadro}
B(T) = B_0 + B_1 T^2
\ee
Then, it is probably possible to fit the  lattice data because $B_1 T^2$ gives the correct $O(1/T^2)$ behavior 
above the peak and at the same time $B_0$ can be adjusted to optimize the curve  below the peak. 

But, as 
$\Delta$ scales with $\simeq N_c(N_c^2-1)$ near $T_c$ and with  $(N_c^2-1)$ for larger temperature, 
the scaling behavior of $B_0$ and $B_1$ with $N_c$ should be different. In particular the
scaling near the critical point can be understood as a violation of the t'Hooft limit
condition $g^2 \simeq 1/N_c$
because
\be
B_0 \simeq   \frac{\beta(g)}{g} <G_{\mu \nu}^a G_{\mu \nu a} > \simeq g^2 \gamma N_c (N_c^2 -1)
\ee
where the constant $\gamma$ contains numerical factors and an average, $N_c$ independent, condensate per gluon. 
Therefore, $\Delta \simeq N_c(N_c^2-1)$ near $T_c$ implies $g^2\neq O(1/N_c)$ as seen before for
the mass term.
However  it must be remarked that there is no indication in lattice computation
\cite{digiacomo} of the $T^2$ behavior in Eq. (\ref{tquadro}) up to $T\simeq 1.5T_c$.

The points discussed above indicate that it is very unlikely that the condensate alone could explain the behavior of 
the interaction measure $\Delta$. On the contrary, it is reasonable to treat the condensate as a small additional 
piece which has to be added to the quasi-particle contribution to $\Delta$, discussed in the previous Sections.
In fact lattice data allow the insertion of a physically acceptable constant $B_0$, with a (minor) role only in the 
region near $T_c$, without qualitatively changing the fit to the data shown in Section III.



\vskip 20pt

\section{V. Comments and Conclusions}

The previous results show that a quasi-particle approach, where the effective mass is related with the
features of the deconfinement transition, gives a very good desciption of the interaction measure and of the 
thermodynamical quantities for weak first order phase transition. 

In our opinion this is not so surprising. Indeed, a quasi-particle approach means that the 
relevant dynamics is contained in the  two-point function and
fluctuactions have a minor role.
A general framework to describe this behavior in quantum field theory is the effective potential for composite
operators  (CJT) \cite{cjt} in the so-called Gaussian approximation. It has been extensively applied at finite
temperature for scalar, fermion and gauge theories and  naturally leads to a first order phase transition 
to restore the symmetries, although this conclusion has to be confirmed by other more reliable methods( $\epsilon$ 
expansion, lattice simulations,etc).

We find that the scaling of the interaction measure with $(N_c^2-1)$ is observed for $T > 1.1 T_c$ and clearly 
violated near $T_c$.
Accordingly, above the peak temperature of the interaction measure,
 the mass behavior is independent on $N_c$ in agreement with a perturbative parametric
dependence on $g^2 \, N_c$ and the $1/\sqrt{N_c}$ t'Hooft scaling of the coupling which, on the other hand, is broken 
very close to $T_c$.
Moreover  by combining this aspect with the almost constant behavior of $\Delta*T^2$ in the range 
$ 1.1 T_c < T < 4 T_c$, 
it turns out difficult, in our opinion, to describe the interaction measure by a temperature dependent bag pressure 
and/or gluon condensate above $T_c$.  An 
interesting point is to find the connection with the confined phase below $T_c$ which can be described by a glueball 
gas plus bag pressure \cite{buisseret}   

It would be also useful to analyze the $SU(2)$ case because there is a second order phase transition , with a 
corresponding divergent effective mass
at the critical point, and a small number  of colors. However the $SU(2)$ data are quite old and ,unfortunately, there 
does not seem to exist any lattice 
study allowing an extrapolation to the continuum, thus eliminating finite  lattice size effects \cite{engels1}.

The meaning of the effective gluon mass as discussed in Sect. II, which is similar to a "colored glue-lump", is 
probably related with the dynamical mechanism of the transition, bubble nucleation for example, and a deeper 
understanding in this direction would be extremely appealing.

Following the same strategy here applied, the next step will be the analysis of full QCD 
to study the role of fermions in 
modifing the effective gluon properties related to  the deconfined phase cross-over.

\noindent
{\bf Acknowledgements}
The authors thank M.P. Lombardo and H. Satz, for useful comments.
This work has been partially supported by the FIRB Research Grant
RBFR0814TT provided by the MIUR.

\end{document}